# Assessment of fiducial motion in CBCT projections of the abdominal tumor using template matching and sequential stereo triangulation

Oluwaseyi M. Oderinde[1,4], Hassan Mostafavi[2], Daniel Simpson[1], James Murphy[1], Gwe-Ya Kim[1], Laura I. Cerviño[1, 3]

[1]*Department of Radiation Medicine and Applied Science, University of California San Diego, La Jolla, CA, 92093;*

[2]*Varian Medical Systems, Palo Alto, CA, USA;*

[3] *Department of Medical Physics, Memorial Sloan Kettering Cancer Center, New York, NY, 10065*

[4]*Department of Physics, University of Witwatersrand, Johannesburg, Gauteng, South Africa*

*Corresponding author @Department of Physics, University of Witwatersrand, Johannesburg, Gauteng, South Africa; Email: Oluwaseyi.oderinde@wits.ac.za*

## Abstract

**Purpose:** To assess the fiducial motion in abdominal stereotactic body radiotherapy (SBRT) using the cone-beam computed tomography (CBCT) projections acquired for pre-treatment patient set-up.

**Materials and Methods:** Pre-treatment CBCT projections, and anterior-posterior (AP) and lateral (LAT) pair of fluoroscopic sequences of 7 pancreatic and 6 liver SBRT patients with implanted fiducials, for a total of 49 treatment fractions, were analyzed retrospectively. A tracking algorithm based on template matching and sequential stereo triangulation algorithms was used to track the fiducials in both the CBCT projections and the fluoro sequence pairs. To measure the accuracy of the tracking algorithm used, the CBCT and fluoro pair projections of a motion phantom with implanted fiducials undergoing a sinusoidal motion were tracked and the results were compared with the known phantom trajectory. To demonstrate the potential of tracking the CBCT projections for pre-treatment setup adjustment, we compared the internal-external correlation coefficient in SI coordinate and $5^{th} - 95^{th}$ percentiles motion range in LAT, SI and AP coordiantes, derived from CBCT tracking with those derived from fluoro pair tracking. Lastly, we predicted the clinicial couch adjustment from CBCT tracking and compared it with the actual clincal couch decision that was made during the patients' treatment.

**Results:** In the phantom study, the tracked SI trajectory agreed with the ground truth with a root mean square error (RMSE) of 0.297, 0.225 and 0.222 mm for CBCT, AP, and LAT projections, respectively. The external-internal motion correlation coefficients (*coeff*) for CBCT, AP, and LAT fluoro projections were 0.94±0.05, 0.97±0.06, and 0.96±0.05, respectively for pancreas patients. In liver patients, the average *coeff* were 0.94±0.07, 0.97±0.03, and 0.97±0.02 for CBCT, AP and LAT fluoro, respectively. In 3D coordinate, the fiducial motion ranges for pancreas cases were 9.90±3.52 mm, 10.65±5.91 mm, and 10.74±6.24 mm for CBCT, AP, and LAT fluoro, respectively, while in liver, they were 13.93±3.39 mm, 11.17±3.75 mm, and 11.52±4.33 mm, respectively. Prediction of couch adjustment in LAT, SI and AP coordinates from CBCT tracking agrees with the actual clinical couch correction within 0.92±0.74 mm, 1.37±1.26 mm, and 0.68±0.56 mm for pancreas cases and within 1.12±0.96 mm, 1.15±0.92 mm and 0.90±0.86 mm for liver cases, respectively.

**Conclusion:** Tracking of pre-treatment CBCT projections using template matching and sequential stereo triangulation is suitable to assess fiducial motion and to adjust the patient setup for abdominal SBRT. CBCT can be used for motion modeling, potentially eliminating the need for the additional fluoroscopic pair acquisition, and thus reducing both the imaging dose to the patient and the total treatment time.

**Keywords:** Fiducial tracking, CBCT, fluoroscopy, template matching, external-internal motion correlation, sequential stereo

1. Introduction

In stereotactic body radiotherapy (SBRT), fiducial markers are frequently implanted near tumors that are subjected to respiratory-induced motion for aiding with patient setup and respiratory gating (1,2). Hypo-fractionated SBRT treatments deliver a high dose to the tumor site. For this reason, treatments need to be greatly monitored and guided in abdominal cases, where internal target positional drift may significantly lead to suboptimal treatment and wider uncertainty margins (3). Respiratory gating using internal fiducials has reduced the motion and improved tumor-to-normal tissue ratio in abdominal cases (4,5). Since the magnitude of internal motion could be greater than 20 mm peak to peak, large treatment margins are not encouraged due to proximity, alongside with low tolerance dose, of the critical structures (6,7).

During simulation of radiotherapy of the tumors that are subjected to respiratory motion, patients are routinely scanned with a 4D-CT to get an accurate characterization of tumor and normal tissue motion for effective treatment planning and optimized dose delivery. Notwithstanding, irregular breathing pattern, day-to-day changes of anatomical structure, and baseline shifts may lead to uncertainty margins after the 4D-CT scan. Therefore, pre-treatment imaging becomes necessary to update the patient setup just before the treatment sessions and to reduce the uncertainty margins.

It is common practice in pre-treatment imaging of liver and pancreas SBRT patients to first acquire CBCT projections, followed by fluoroscopic imaging. Fluoroscopic imaging, which provides a high-resolution and real-time image information, is routinely used to complement the blurred images of tumor/markers in the average 3D-CBCT images (8). However, accurate tracking of the tumor/fiducials in CBCT projections has the potential to obviate the routine use of fluoroscopic imaging, thus saving the total treatment time and reducing the pre-treatment imaging dose.

Previous studies on abdominal RT have shown the good soft-tissue contrast provided by CBCT and the improved treatment setup errors using CBCT matching on fiducials rather than on bony anatomy (9–11). CBCT imaging acquisition time is limited by gantry rotation speed, 630 projections at full standard $360^0$ CBCT half-fan scan can be acquired within 60 seconds with about 2cGy patient dose, approximately (12). Recently, there have been feasibility studies on the tracking of internal tumor/fiducial motion in CBCT projections using various algorithms (13–15).

In this study we use template matching for fiducial tracking. The algorithm is part of an offline research tool (RapidTrack, Varian Medical Systems) designed to track target motion in pre-treatment and intrafraction images acquired by an onboard imager. This study centered on achieving the following: (1) evaluating the tracking accuracy of RapidTrack algorithm, (2) evaluating the correlation coefficient of the internal target motion with an external breathing surrogate signal (RPM, Varian Medical Systems), (3) assessing the abdominal tumor motion range from pancreas and liver cases during SBRT, and (4) prediction of clinical couch adjustment using CBCT projections.

## 2. Materials and Methods

In this study we first evaluated the accuracy of RapidTrack template matching algorithm using a motion phantom with known motion. We then investigated the use of CBCT projections for internal marker tracking and pre-treatment patient setup using pancreas and liver patient data,. This first section describes the template matching process. This is followed by a description of the imaging and data analysis methods for both phantom and patient studies.

### 2.1. Fiducial marker tracking

A software suite for offline research, RapidTrack, which includes template matching and sequential stereo triangulation was utilized in this study. The general procedure is shown in **Figure 1**.

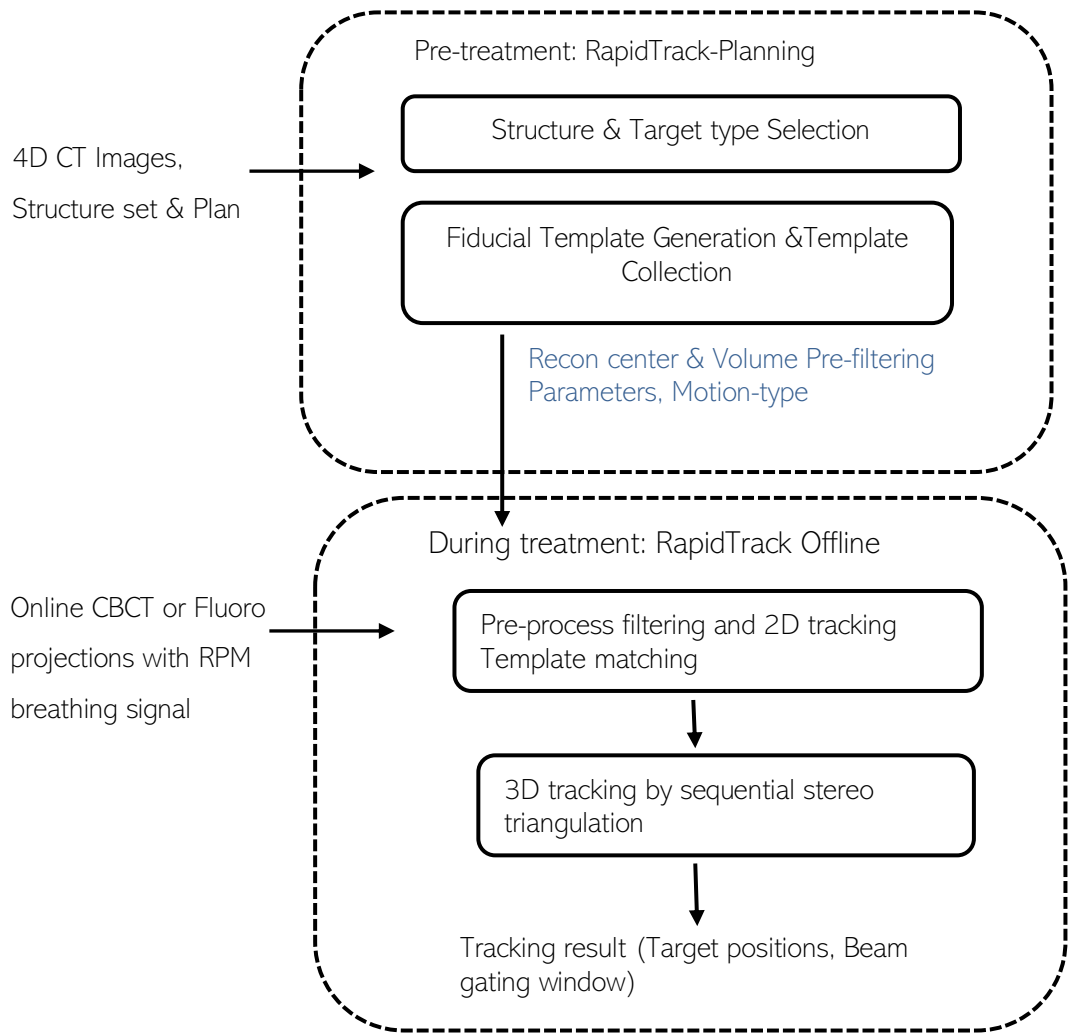

*Fig. 1*: *Workflow for fiducial tracking with abdominal CBCT and fluoroscopic projections.*

For template generation, treatment plan data were exported from the Record and Verify information system. The exported data contains the CT images alongside with the RT structures and RT plan. These data were imported in the template generation tool. The algorithm constructs a volume of interest (VOI) around the selected structures (fiducial markers) to create a collection of templates by forward projection, similar to digitally reconstructed radiographs (DRRs), for all imaging directions at 1-degree spacing. In each projection the template with the corresponding direction is matched to a 2D search window after applying an optimizing pre-filter, thus estimating the pixel-domain target offset from the plan position for that projection. Examples of template matching results are shown in **Figure 2**. The collection of these pixel domain offsets and the associated X-ray source/detector geometries form the input to the sequential stereo triangulation from which 3D target tracks are calculated.

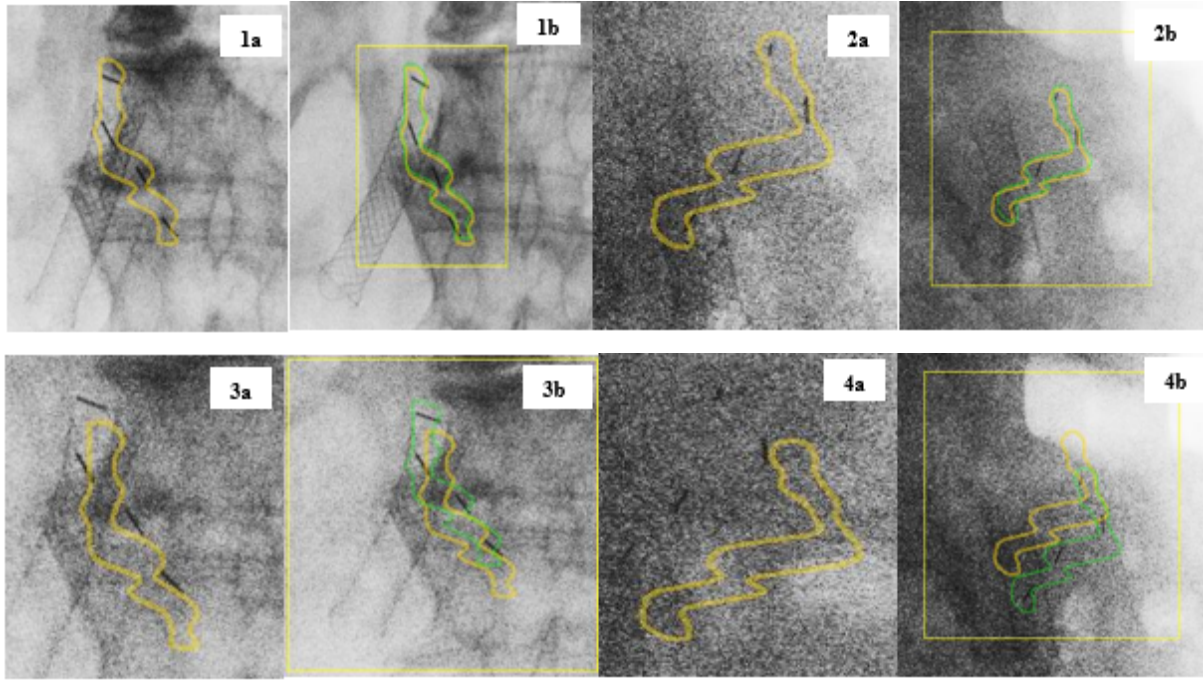

*Fig. 2*: Example of tracked fluoroscopic (top) and CBCT (bottom) projections using template matching + sequential stereo triangulation. 1a & 3a) before tracking for fluoro & CBCT at $0^0$; 1b & 3b) after tracking for fluoro & CBCT at $0^0$; 2a & 4a) before tracking for fluoro & CBCT at $270^0$; and 2b & 4b) after tracking for fluoro & CBCT at $270^0$ (yellow contour = target reference position, green contour = target tracked position)

By tracking the CBCT projections (~900 projections at 0.4 degree spacing acquired over 360 degree rotation) the above process generates the 3D track of the target using the monoscopic onboard kV imager of the LINAC. In order to achieve higher template matching accuracy the lateral projections of CBCT, which tend to have lower contrast-to-noise ratio, were exluded. The used ranges of CBCT projection angles were typically ~$120^0$-$280^0$ and ~$330^0$-$30^0$. Fluoro projections were tracked by template matching similar to CBCT projections. The 2D pixel-domain tracks of the AP fluoro sequence were combined with the LAT fluoro sequence to form the input to the sequential stereo calculation of 3D tracks.

**2.2. Phantom study**

A QUASAR™ multi-purpose acrylic body phantom and respiratory motion platform (Modus Medical Devices Inc., London; Ontario, Canada) were used for the evaluation of fiducial tracking with RapidTrack. The acrylic oval-shaped phantom measures 30 cm wide, 20 cm high, and 12 cm long. A sinusoidal motion with a nominal 10 mm amplitude and 4 second period was fed to the motion platform, which was set up to simulate SI motion. Three gold markers (MTNW887808, CIVCO Medical Solutions, Kalona, IA, USA) were inserted into the middle ion-chamber placement hole of the phantom. The cylindrical markers were 3 mm long and 0.4 mm in diameter, and were spaced approximately 1 cm apart. The motion platform moved in the SI direction, while the external motion signal was generated by tracking the synchronized anterior-posterior (AP) motion of the RPM marker block on the phantom platform simulating the chest wall, as shown in **Figure 3**.

4DCT scan of the phantom in supine position was acquired with a 64-slice CT scanner (GE LightSpeed, Ge Healthcare, Pasadena, CA, USA). All the 4DCT scans were reconstructed for ten breathing phases in 10% steps to mimic the clinical protocol. The templates were generated from the planning 4DCT by combining the $30^0$ to $70^0$ phases. CBCT and fluoroscopic projections of the phantom were acquired on the TrueBeam linear accelerator while simultaneously acquiring the external RPM signal.

The scanning geometry used in phantom experiment was similar to the clinical process used for the study patients with SAD=100 cm, SDD = 150 cm and pixel size of 0.388 × 0.388 mm, half-fan full rotation acquisition for CBCT at 6 deg/sec and ~15 fps, and fixed gantry AP and LAT fluoro acquisition at ~7 fps. The kV technique were 125 kVP, 15 mA, 20 ms for CBCT, and 80 kVP, 57 mA, 18 ms for fluoro.

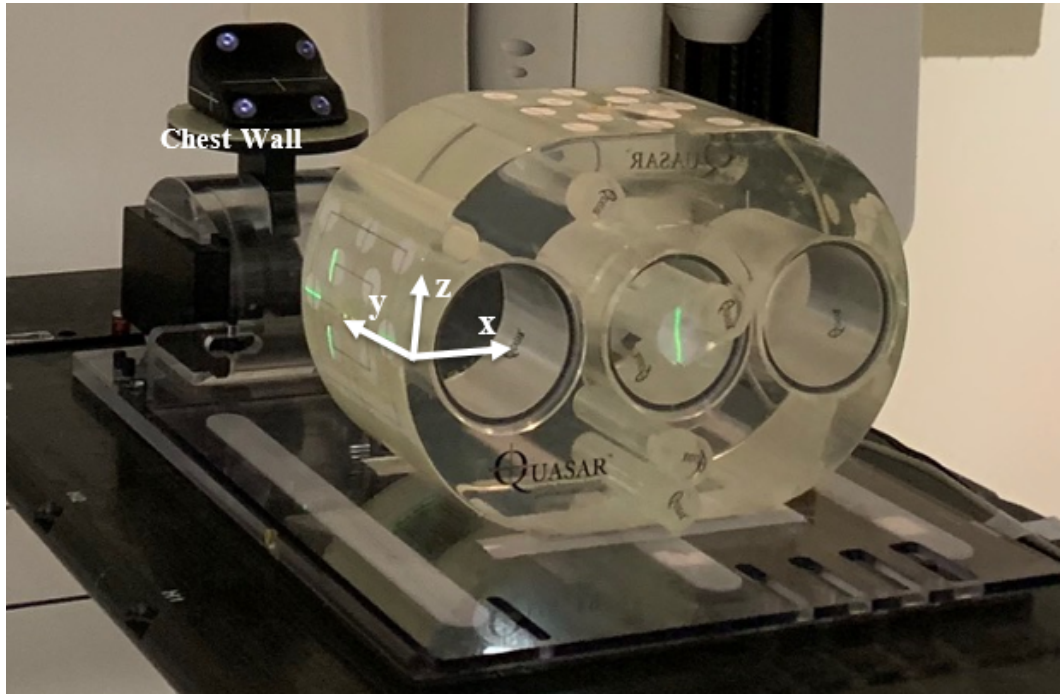

***Fig. 3***: *QUASAR<sup>TM</sup> multi-purpose body phantom and respiratory motion platform during the 4D-CT scan. x-, y- and z-axes are the lateral (left-right), superior-inferior, and anterior-posterior direction, respectively.*

### 2.3. Patient study

At the University of California San Diego, data from three pancreatic and three liver cancer patients treated with gated stereotactic/hypofractionated irradiation during free-breathing was retrospectively analyzed for this study under a protocol approved by the Institutional Review Board. These are patients with four cylindrical-shaped gold fiducial markers implanted in or near the tumor. Pre-treatment CBCT and fluoro projections of a total of 49 treatment fractions were analyzed.

The patients were scanned at free breathing in the supine position, and the RPM device was utilized to monitor the breathing motion for the 4DCT acquisition. All the 4DCT scans were built from ten breathing phases in 10% steps. For template generation purposes, the combined CT image corresponding to 30% to 70% the breathing phases, which corresponds to the end-of-exhale gating window, were used. The implanted fiducials were contoured with an extra 1 mm margin using maximum-intensity-projection CT ($CT_{MIP}$) images.

In each treatment fraction the patients were initially repositioned by orthogonal kV/kV imaging based on bone matching, and then further adjusted by matching the fiducials using the CBCT. Lastly, visualization of respiratory-induced fiducial motion within the gating window was used to refine patient's position using the fluoroscopic images.

### 2.4. Analysis

To validate the RapidTrack algorithm, root-mean square error (RMSE) metric was used to quantify the difference between the tracking result and the ground truth. In addition, the correlation between the RPM signal and the internal SI component of fiducial motion was quantified for CBCT and fluoroscopic projections using the squared Pearson correlation coefficient ($R^2$). We used the $5^{th}$ – $95^{th}$ percentile to deduce the displacement motion range in LAT, SI and AP coordinates.

## 3. Results

### 3.1. Phantom study

**Figure 4** shows the SI component of the tracked fiducial motion. This tracked motion agrees with the sinusoidal ground truth motion to within a root mean square error (RMSE) of 0.297, 0.225 and 0.222 mm for CBCT, AP and LAT fluoroscopic projections, respectively. The calculation of these differences took into account the couch shift that was applied in between CBCT and AP-LAT fluoro pair acquisitions.

The correlation coefficients between the RPM signal and the SI component of motion were 0.995±0.408, 0.997±0.415, and 0.996±0.438 for CBCT and AP-LAT fluoro projections, respectively.

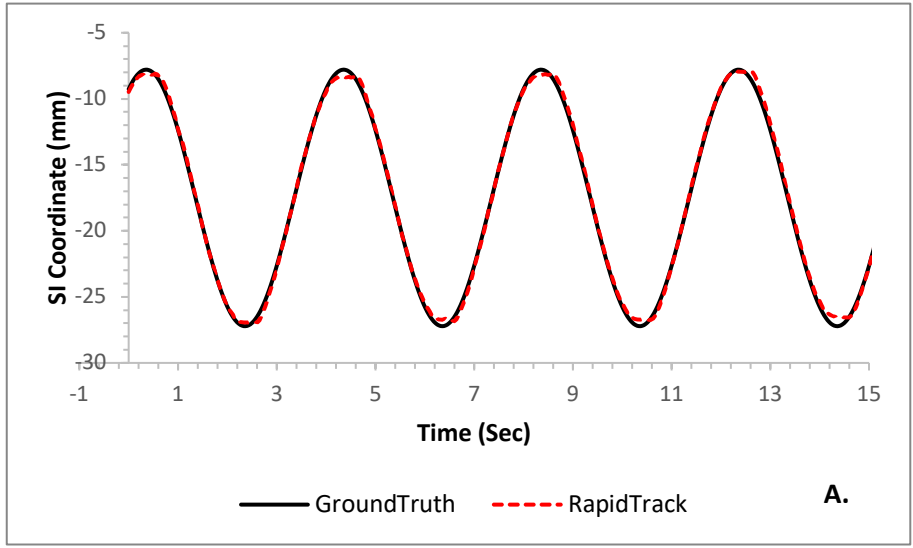

A.

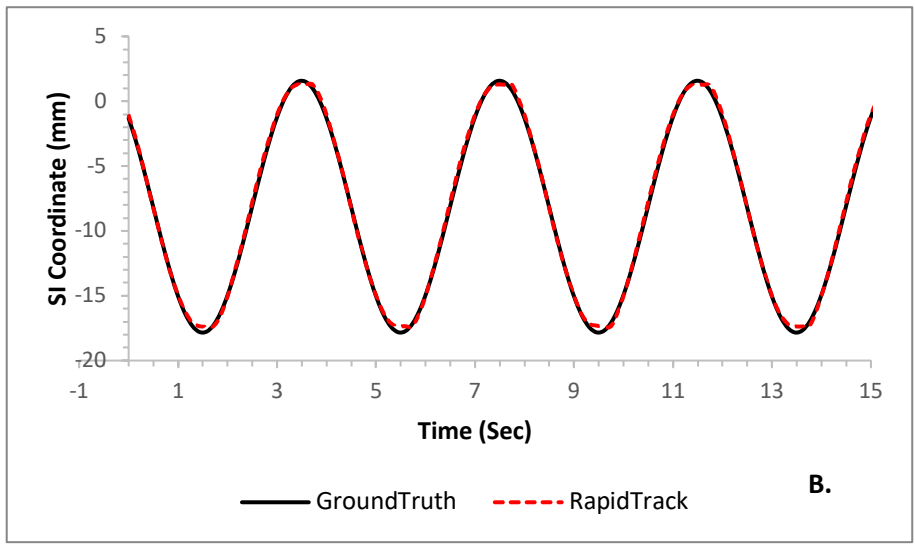

B.

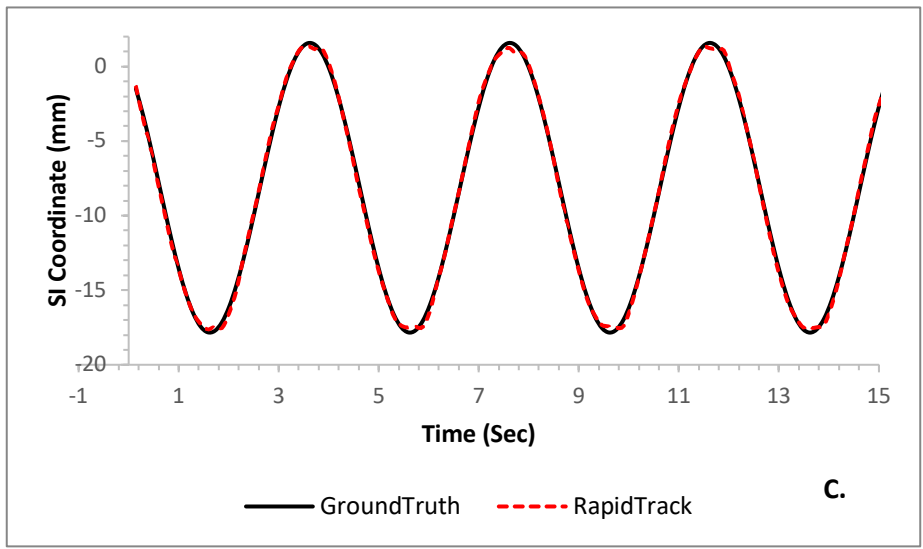

C.

*Fig. 4: Simulated fiducial real-time tracking result in SI coordinate for A) CBCT, B) AP fluoro, and C) LAT fluoro projections*

### 3.2. Patient study

*Pancreas cases*

In pancreas cases, external-internal *coeff* between the RPM signal and the SI component of motion for CBCT, AP, and LAT fluoro projections are shown in **Figure 5**. In pancreas patient two (P2), LAT fluoroscopic projections were missing for treatment fractions two (F2) and four (F4). The average external-internal *coeff* averaged over all patients and fractions for CBCT and AP-LAT fluoro projections were 0.94±0.05, 0.97±0.06, and 0.96±0.05, respectively.

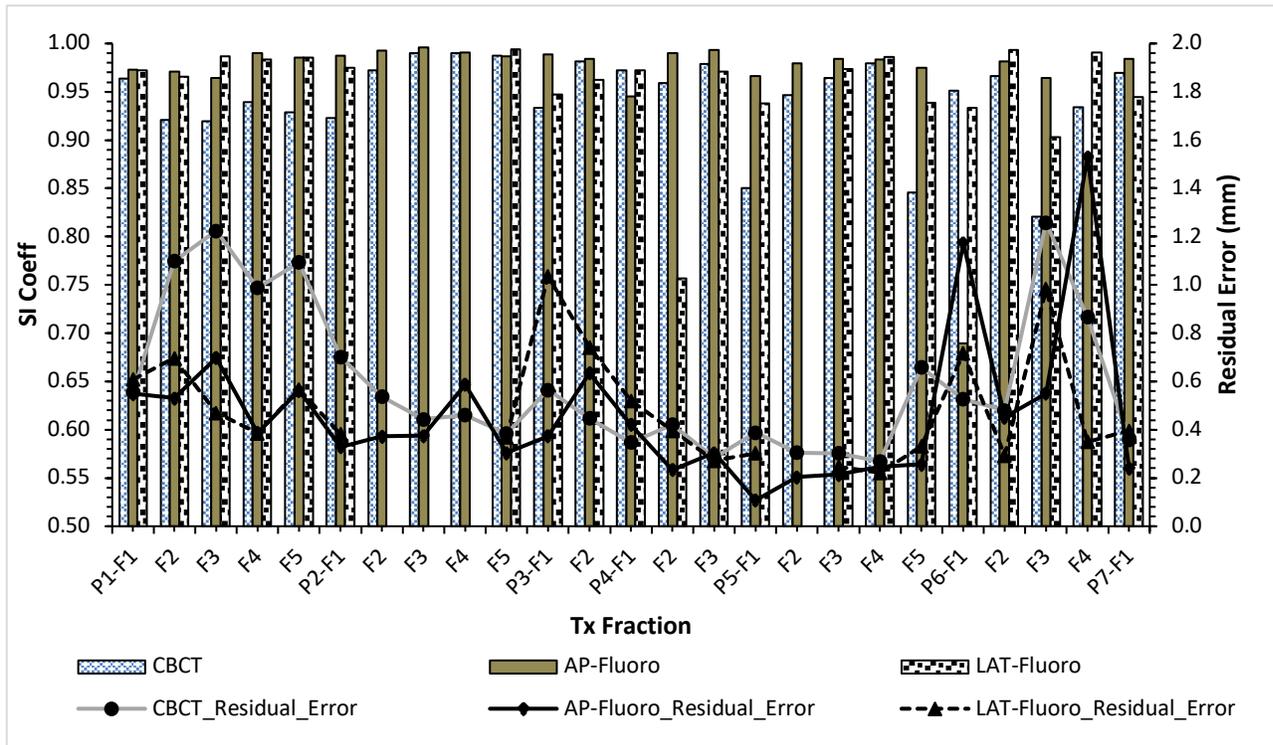

*Fig. 5: Internal-external correlation coefficients for CBCT and fluoroscopic pair projections for pancreatic data*

**Figure 6** shows the 3D tracking results for pancreas fiducial motion using template matching plus sequential stereo to obtain the LR, SI, and AP motion components. In CBCT tracking, the estimated motion range averaged over all patients and fractions were 2.20±1.15 mm, 8.34±3.17 mm, and 4.58±1.93 mm for LAT, SI, and AP coordinates, respectively. In AP fluoro tracking (using LAT as prior 2D tracks), these motion range averages were 3.24±2.82 mm, 7.99±3.35 mm, and 5.29±5.23 mm for LAT, SI, and AP coordinates, respectively. In LAT fluoro tracking (using AP as prior 2D tracks), these results were 3.21±3.78 mm, 7.99±3.43 mm, and 5.34±5.11 mm, respectively. In 3D, the estimated fiducial motion ranges were 9.90±3.52 mm, 10.65±5.91 mm, and 10.74±6.24 mm for CBCT, AP, and LAT fluoroscopic projections, respectively. Using the CBCT images, RapidTrack algorithm predicted the optimized patient setup which agrees with the clinical patient setup that was applied during the SBRT treatment. **Table 1** shows the setups agreed within 0.92±0.74 mm, 1.37±1.26 mm, and 0.68±0.56 mm for LAT, SI and AP coordinates, respectively.

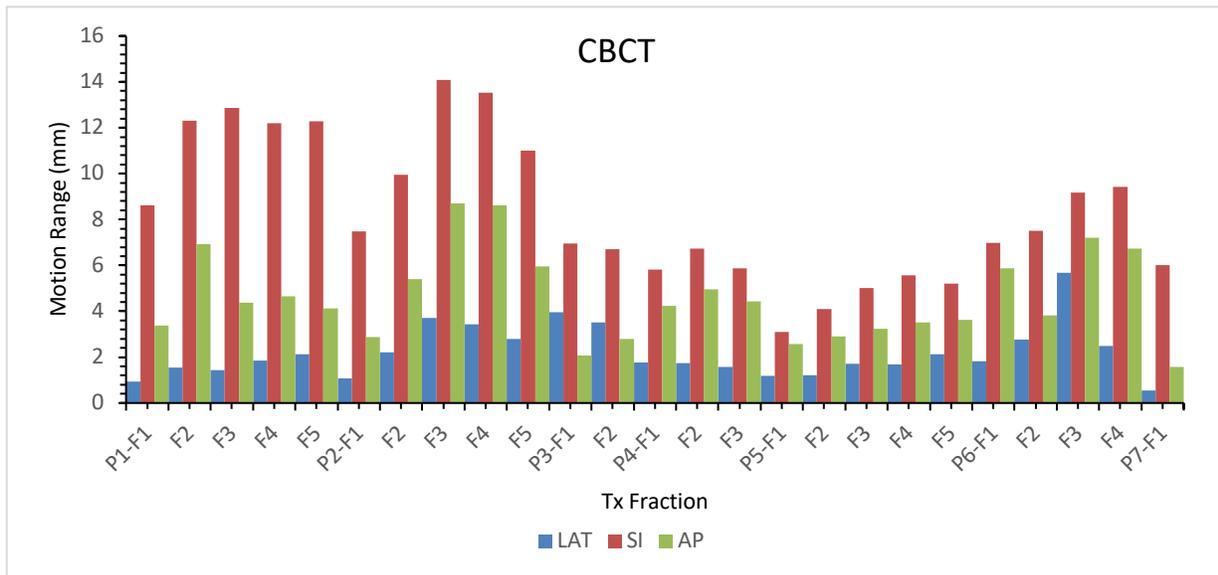

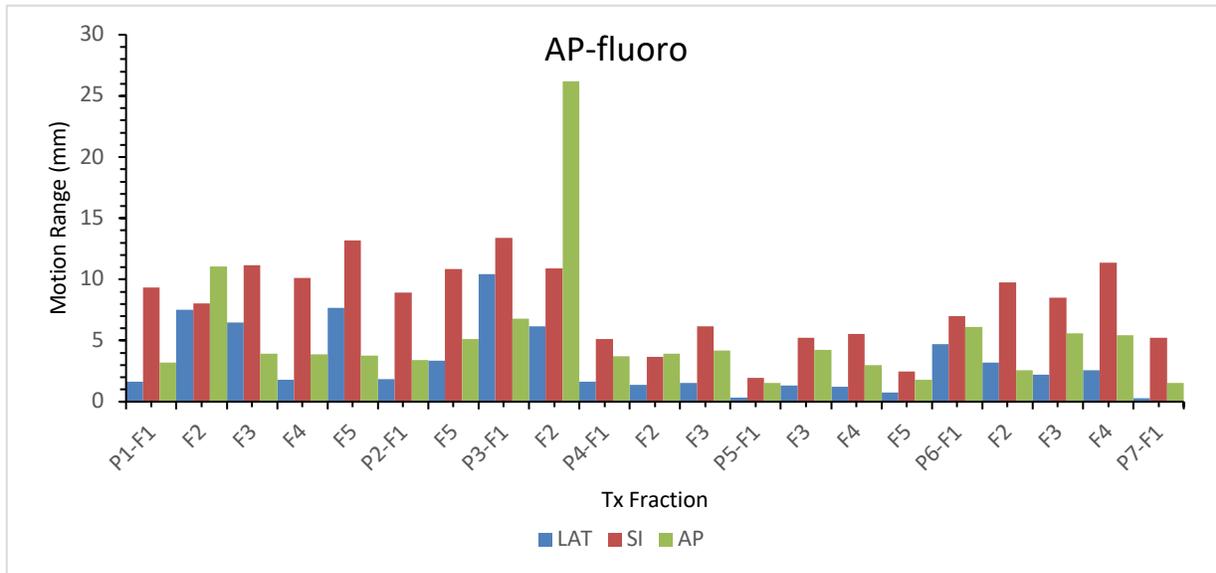

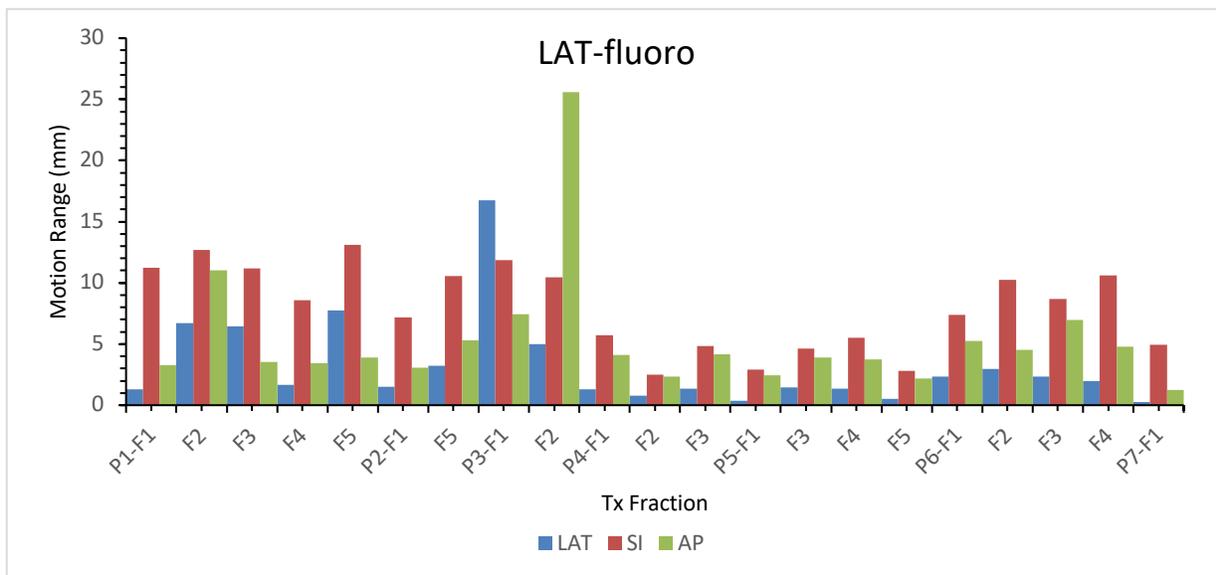

***Fig. 6***: *LAT, SI and AP fiducial motion range in pancreas patients for CBCT and pair (AP and LAT) fluoroscopic projections (P: Patient, C: CBCT projection, A: AP and L: LAT fluoroscopic projection, F: treatment fraction)*

*Table 1: Predicted couch adjustment in LAT, SI and AP direction, which agrees with clinical couch correction for pancreas cases*

| Patient | LAT (mm) | SI (mm) | AP (mm) |
| --- | --- | --- | --- |
| P1 | 0.75±0.30 | 0.94±0.52 | 0.50±0.23 |
| P2 | 1.07±0.88 | 0.88±0.58 | 0.58±0.40 |
| P3 | 1.80±0.00 | 0.65±0.64 | 1.05±1.06 |
| P4 | 0.17±0.21 | 3.13±2.66 | 1.17±1.24 |
| P5 | 0.90±0.73 | 1.66±0.67 | 0.48±0.35 |
| P6 | 1.30±0.39 | 1.15±1.44 | 0.80±0.39 |
| P7 | 0.00±0.00 | 1.50±0.00 | 0.40±0.00 |

*Liver cases*

**Figure 7** shows the external-internal correlation *coeff* between the RPM signal and the SI component of internal target motion for CBCT and AP, LAT fluoro pair. The correlation *coeff* averaged over all patients and fractions for CBCT and and AP, LAT fluro pair were 0.94±0.07, 0.97±0.03, and 0.97±0.02, respectively.

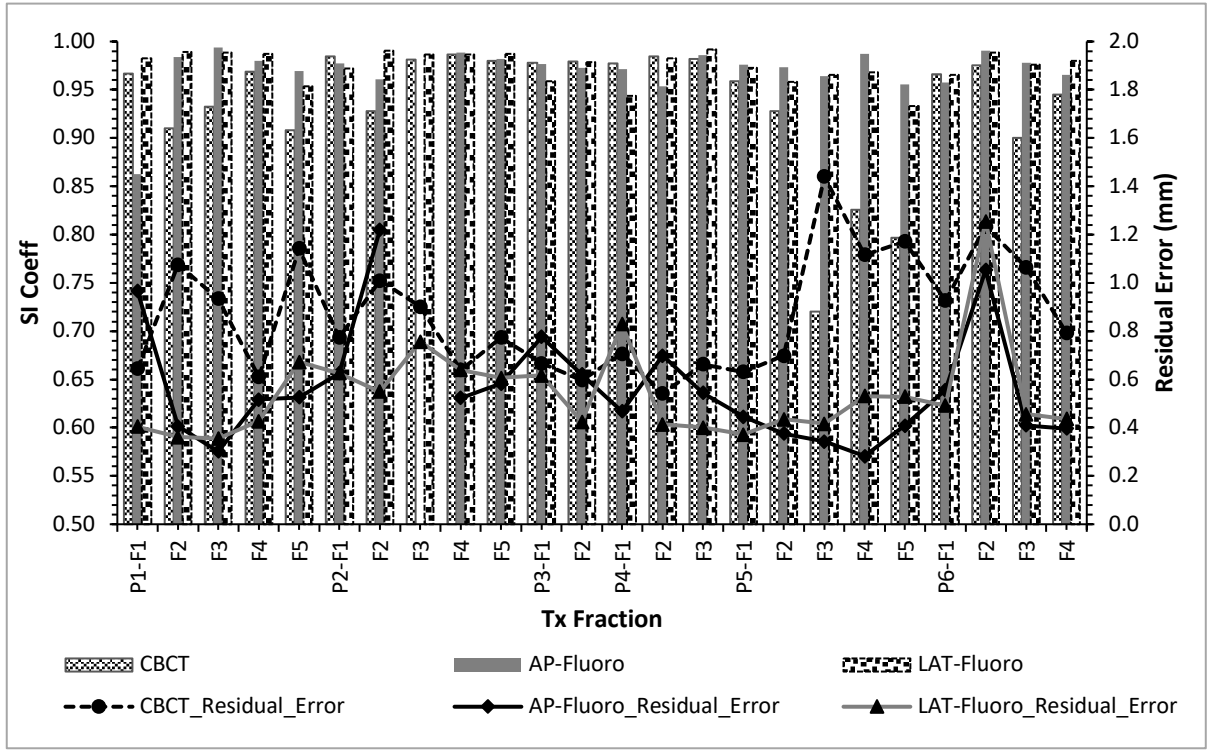

*Fig. 7: Internal-external coeff for CBCT and pair fluoroscopic projections for liver data*

**Figure 8** shows the 3D tracked fiducial motion range in LR, SI, and AP directions for liver treatment fractions. In LR/ SI/ AP coordinates, the motion range averaged over all patients and fractions were 3.56±2.61 mm/12.51±3.83 mm/3.84±0.83 mm for CBCT, 3.17±1.98 mm, 10.04±3.86 mm/2.94±0.88 mm for AP fluoro, and 3.30±2.38 mm/9.98±3.33 mm/3.67±3.33 mmfor LAT fluoro, respectively. Also, the estimated 3D motion ranges were 13.93±3.39 mm, 11.17±3.75 mm, and 11.52±4.33 mm for CBCT, AP, and LAT fluoroscopic projections, respectively. RapidTrack predicted couch adjustment using CBCT images agreed with the clinical corrected couch setup within 1.12±0.96 mm, 1.15±0.92 mm and 0.90±0.86 mm in LAT, SI, and AP coordinates, respectively as shown in **Table 2**.

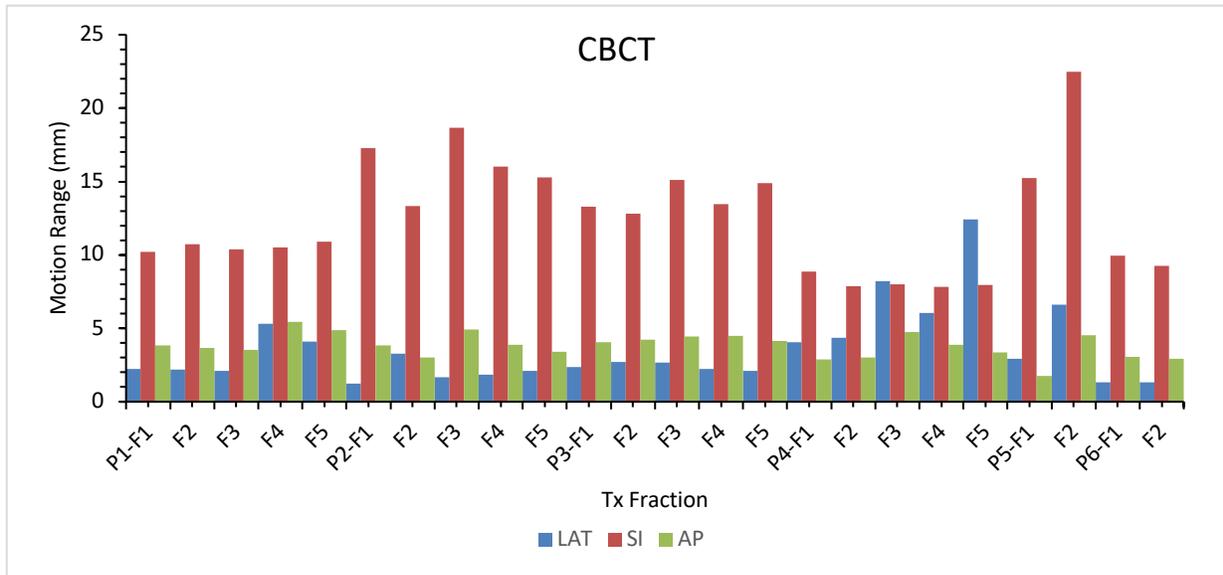

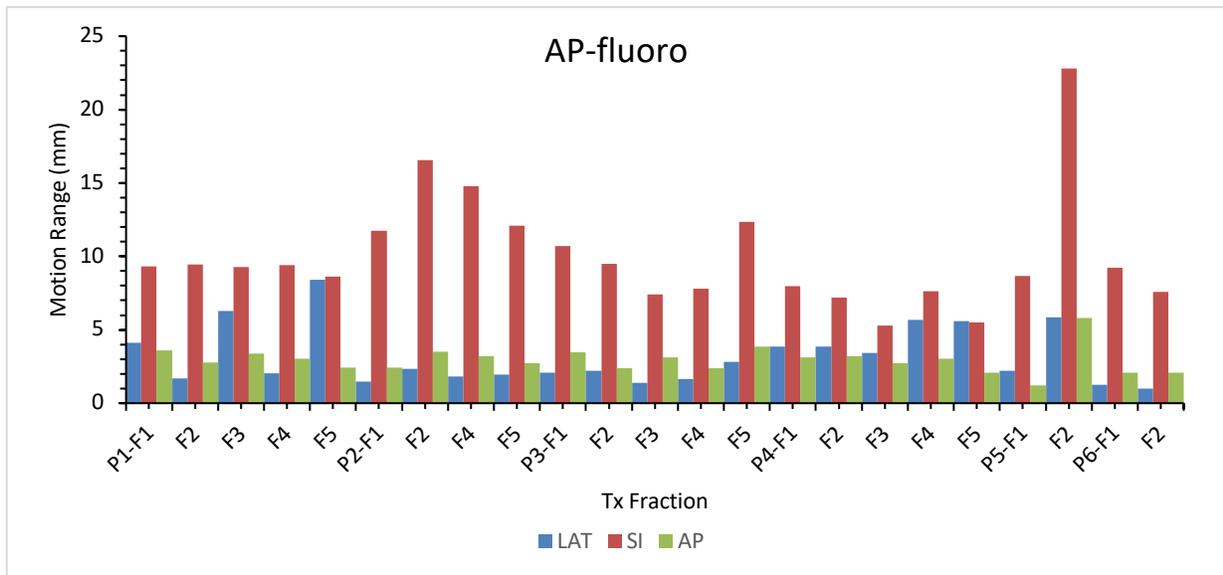

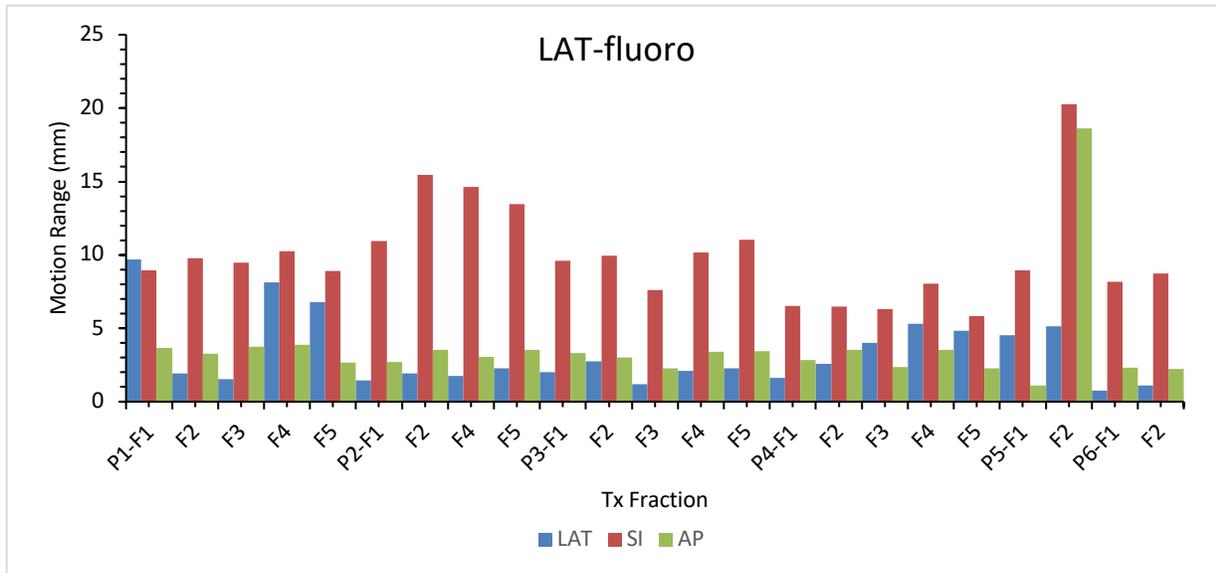

*Fig. 8*: LAT, SI, and AP fiducial motion range in pancreas and Liver patients for CBCT and pair (AP and LAT) fluoroscopic projections (P: Patient, C: CBCT projection, A: AP and L: LAT fluoroscopic projection, F: treatment fraction)

*Table 2*: Predicted couch adjustment in LAT, SI and AP direction, which agrees with clinical couch correction for liver cases

| Patient | LAT (mm) | SI (mm) | AP (mm) |
| --- | --- | --- | --- |
| **P1** | 1.94±1.47 | 0.96±0.63 | 1.34±1.02 |
| **P2** | 0.88±0.77 | 0.58±0.50 | 0.46±0.50 |
| **P3** | 0.62±0.33 | 0.86±0.61 | 0.50±0.31 |
| **P4** | 1.50±0.83 | 1.90±0.80 | 1.60±1.09 |
| **P5** | 0.70±0.14 | 0.65±0.07 | 0.20±0.00 |
| **P6** | 0.40±0.28 | 2.35±2.05 | 0.85±0.78 |

## 4. Discussion

This study has shown that template matching and sequential stereo triangulation can be used to measure the respiratory motion trajectory and range by tracking the fiducials in the pre-treatment CBCT projections and fluoro pair sequences. As a result, the correlation model between an external surrogate, in this case the RPM signal, and the internal motion can be established. In phantom, we assessed the accuracy of the CBCT and fluoro pair tracked positions with the known phantom motion. In patients, we assessed the equivalence of CBCT and fluoro pair tracking results which is conventionally used for pre-treatment patient setup for abdominal tumor SBRT using respiratory and predicted the couch correction for treatment delivery using the CBCT pre-treatment images. The result presented here suggest the use of CBCT projections for clinical couch adjustment in real-time treatment.

Our phantom study shows that the CBCT tracking algorithm is accurate in terms of real-time tracking and motion range calculation. Relatively, we anticipate having less accuracy in patient's study because of scattering and rotational effect in the inhomogeneous body. Also, patients' irregular breathing pattern and possibly an inherent error from imaging modalities could have reduced the accuracy of CBCT correlation coefficient. Furthermore, longer acquisition time for CBCT imaging (≈80.20 sec), which is more than fluoroscopic projection (≈14.38 sec), could have contributed to *coeff* < 0.9. It is logical to deduce that the longer CBCT acquisition time can accommodate patients' irregular breathing in the projections. Notwithstanding, CBCT projections shows the potential to be used for pretreatment patient setup optimization. These projections generated good correlation coefficients just like the fluoroscopic pair projections.

CBCT and fluoro pair were acquired in free-breathing patient treatments, where the breathing amplitude will likely change within a few minutes (16). In our study, we had an average of 8.0 minutes time gap between CBCT and fluoroscopic projections.Therefore, variations within the estimated fiducial motion range using CBCT and Fluoro pair projections are expected. In addition, RapidTrack algorithm generated the 3D fiducial motion ranges using an interdimensional mathematical correlation model. Our study has shown that the largest motion magnitude in pancreas (8.34±3.17 mm) and liver (12.51±3.83 mm) lesions were observed in the SI directions, which follows related studies that show average motion magnitudes between 0.9 and 28.8 mm (17–20).

Conventionally, CBCT has been in use for daily pre-treatment setup, especially for soft tissue matching (8,9), and it can be utilized for respiratory induced tumor/fiducial tracking. Yang J. et al. (21) considered the diaphragm as a good surrogate for abdominal motion, and Wei and Chao (13) extracted diaphragm motion from abdominal CBCT projections using linear regression optimization algorithm. The study evaluated the average tracking error, which is higher than fluoroscopic imaging. However, diaphragm tracking might not the best surrogate, since there are uncertainties in phase relationships between the tumor and the anatomical structure, and interclass correlation coefficient, especially in LAT direction (21,22). Also, Jang et al. (14) article on evaluating the use of CBCT projection to predict daily motion was limited to phantom study. Rankine et al. (15) developed a dynamic programming algorithm for patient set-up by tracking the optical marker position from CBCT projection. This latter study evaluated the motion range of the tracked positions and concluded that CBCT could accurately predict intrafractional tumor motion. Our study predicts internal motion margin using template matching algorithm and sequential stereo-triangulation technique, which was designed for both retrospective and prospective patient pre-treatment setup. In addition, during the pre-filtering process, RapidTrack algorithm can filter the low-quality images (weak intensity gradient and loud interference/noisy frames) automatically. These blurred projections (low-signal-to-noise ratio) are caused by gantry rotation alongside with lateral projections in patients (due to the presence of ribs and bony structures), which can probably skew the tracking results detected within the search window.

Furthermore, optimization of patient setup for treatment delivery is manually achieved based on the clinical team expertise. In our study, the RapidTrack algorithm automatically predicted the optimized couch setups using CBCT images, which agree (within 1.37±1.26 mm and 1.15±0.92 mm for pancreas and liver cases in SI coordinate, respectively) with the actual clinical couch decisions that were made for the treatment cases. Automation of the clinical couch correction could help in reducing the total treatment time, saving the necessary time for the clinical team to re-check and be sure of making the best decision for each fraction. Hence, clinical team's expertise in making this decision may not be crucial since the algorithm can serve as a validation tool.

RapidTrack tool is a user-friendly software used for the trajectory of the day measurement, patient setup optimization, and intrafraction positioning monitoring. Since the CBCT projections has shown the potential to be used for pre-treatment setup, the routine use of fluoro pair projections for clinical couch shift decision can soon become obsolete as we look forward to implementing our pre-treatment imaging technique for abdominal SBRT cases with real-time tracking algorithm. However, the decision not to routinely use both CBCT and fluoro pair still rely on the patient care team and can be influenced based on individual patient's need.

## 5. Conclusion

Taking advantage of template matching and sequential stereo triangulation, CBCT projections can assess fiducial motion trajectory, in lieu of using fluoroscopic projections, potentially to the benefit of patients in terms of reducing treatment time and imaging dose.

## 6. Declaration of conflict of interest

*This project was funded by Varian Medical Systems, Palo Alto, CA, USA*

22 | P a g e

during treatment. Am J Clin Oncol Cancer Clin Trials. 2009;32(4):364–8.